\newcommand{\beq}{\begin{equation}}
\newcommand{\enq}{\end{equation}}
\newcommand{\bee}{\begin{eqnarray}}
\newcommand{\ene}{\end{eqnarray}}
\newcommand{\bem}{\begin{mathletters}}
\newcommand{\enm}{\end{mathletters}}

\documentstyle[eqsecnum,aps]{revtex}

\begin{document}
\draft

\title{``QUADRATIC SOLITONS" IN NONLINEAR DYNAMICAL LATTICES}
\author{Vladimir V. Konotop$^1$ and Boris A. Malomed$^2$}
\address{
$^1$ Centro de F\'{\i}sica da Materia Condensada,\\
Complexo Interdisciplinar da Universidade de Lisboa, Av. \\
Prof. Gama Pinto, 2, Lisbon, P-1649-003 Portugal  \\ and 
Departamento de F\'{\i}sica, Faculdade de Ci\^encias, Universidade de \\
Lisboa, Campo Grande, Lisbon, P-1749-016 Portugal \\  
$^2$ 
Department of Interdisciplinary Studies,\\ 
Faculty of Engineering, 
Tel Aviv University,\\
Tel Aviv 69978, Israel}
\date{\today}
\maketitle

\begin{abstract}
We propose a lattice model, in both one- and multidimensional versions,
which may give rise to matching conditions necessary for the generation of
solitons through the second-harmonic generation. The model describes an
array of linearly coupled two-component dipoles in an anisotropic nonlinear
host medium. Unlike this discrete system, its continuum counterpart gives
rise to the matching conditions only in a trivial degenerate situation. A
system of nonlinear evolution equations for slowly varying envelope
functions of the resonantly coupled fundamental- and second-harmonic waves
is derived. In the one-dimensional case, it coincides with the standard
system known in nonlinear optics, which gives rise to stable solitons. In
the multidimensional case, the system prove to be more general than its
counterpart in optics, because of the anisotropy of the underlying lattice
model.
\end{abstract}

\pacs{PACS numbers: }


\narrowtext

\section{Introduction and Formulation of the Model}

Solitons supported by quadratic nonlinearities, frequently called
``quadratic solitons'' (QS), have recently attracted a great deal of
attention in nonlinear optics, where they can be easily observed
experimentally in the spatial domain (as self-supported cylindrical beams in
a bulk medium \cite{exp1}, or stripes in planar waveguides \cite{exp2}). The
experimental observation of QS in the temporal domain \cite{DiTrapani}, and
of spatiotemporal QS, or ``light bullets'', in three-dimensional samples 
\cite{Frank}, has been recently reported too, although it proved to be much
more difficult to observe temporal solitons than the spatial ones.

Besides optics, quadratic nonlinearities coexisting with linear dispersion
(which is necessary to support solitons in balance with the nonlinearity)
occur in many other physical media, which suggests a possibility of
theoretical prediction and experimental observation of QS in systems other
than optical. In particular, it is expected that the experimental
observation of QS in the temporal and spatiotemporal domains, which is
fairly difficult in the optical media, could be easier in other physical
settings. In line with this idea, temporal and spatiotemporal QS have been
recently predicted in a system of resonantly coupled internal waves in an
inviscid stratified fluid \cite{Roger}.

An objective of the present work is to predict QS in dynamical lattices
featuring quadratic nonlinearity. The resonant second harmonic generation
(SHG) in nonlinear lattices, as well as more general resonant three-wave
interactions, have been described in Refs. \cite{VK,VK2}. However, the
scaling adopted in those papers did not imply existence of QS. In order to
obtain them, one needs to include into consideration the dispersion (which
is by itself an intrinsic property of a discrete system).

We will demonstrate that any generic quadratic nonlinearity can easily give
rise to a soliton, provided that the model's linear part satisfies necessary
matching conditions: in the lowest-order approximation, phase and group
velocities of the fundamental and second harmonics (to be abbreviated as FH
and SH, respectively) must coincide at some value of the wavenumber. It
turns out that these two conditions are difficult to satisfy in some
standard dynamical lattices. For instance, one can check that the matching
never takes place in monoatomic lattices, while in a diatomic lattice, whose
linear spectrum consists of two branches, the matching occurs only in two
cases (see, e.g., \cite{VK2}), when the FH wavenumber is exactly at the edge
or center of the first Brillouin zone (BZ). In either case, the coincidence
of the FH and SH group velocities is trivially provided by the fact that
both are equal to zero. Even this trivial matching in the diatomic lattices
can give rise to QS with a zero or small velocity. However, our aim here is
to propose a physically realistic model, describing a chain of two-component
dipoles in an anisotropic host medium, that readily meets the matching
conditions in a nontrivial way (while a continuum limit of the model can do
this only in a trivial degenerate situation). We argue that the system may,
in fact, be a representative of a wider class of models (e.g., those
describing long molecular chains), where solitons can be generated by a
nontrivial resonance between FH and SH.

In section 2 we formulate the model and produce results of its linear
analysis. The nonlinear equations for the slowly varying amplitudes of the
resonantly interacting FH and SH waves are derived in section 3. A
generalization to the two- and three-dimensional cases are briefly
considered in section 4. In the one-dimensional case, the equations turn out
to be essentially equivalent to the standard SHG equations in nonlinear
optics that are well known to give rise to stable QS (see, e.g., Ref. \cite
{BurKiv}), while in the multidimensional case they may be more general than
those known in optics \cite{bullet}. The paper is concluded by section 4.

\section{The Model and Its Linear Analysis}

We consider a chain of dipoles with two components $x_n(t)$ and $y_n(t)$
immersed into a nonlinear anisotropic host medium. Physically, the dipoles
may represent, e.g., bound electrons in a chain of impurity atoms, or in a
string of atoms which is a part of the host medium.

Because of the anisotropy of the system, the intersite coupling constants $%
\kappa _{x,y}$ of each of two sublattices, $x_{n}(t)$ and $y_{n}(t)$, may be
different. As usual, we assume that the medium's fundamental symmetry group
lacks inversion invariance, hence terms quadratic in the dynamical variables
may appear in the equations of motion. Thus, the model takes the form 
\begin{mathletters}
\begin{eqnarray}
&&\ddot{x}_{n}+\Omega _{x}^{2}x_{n}-\kappa _{x}\left(
x_{n+1}-2x_{n}+x_{n-1}\right)   \nonumber \\  \label{eq1a}
&=&a_{xx}x_{n}^{2}+a_{xy}x_{n}y_{n}+a_{yy}y_{n}^{2}\,, \\
&&\ddot{y}_{n}+\Omega _{y}^{2}y_{n}-\kappa _{y}\left(
y_{n+1}-2y_{n}+y_{n-1}\right)   \nonumber \\ \label{eq1b}
&=&b_{xx}x_{n}^{2}+b_{xy}x_{n}y_{n}+b_{yy}y_{n}^{2}\,,
\end{eqnarray}
where $\Omega _{x,y}$ are the frequencies generated by the onsite potential,
the coefficients $a$ and $b$ describe the onsite nonlinearity, and the
effective masses of the two components do not appear in the equations as we
can divide by them and accordingly renormalize the other coefficients. The
form in which these equations are written implies that lattice spacing is
set to be equal to one.

Note that cubic nonlinear terms, that were omitted in Eqs. (1) and (2) with
the only purpose to render the equations shorter, can be readily added. It
is well known that adding the cubic terms do not, generally, destroy QS \cite
{BurKivTr,embedded}. It is also relevant to note that, although we wrote
only onsite nonlinearities in Eqs. (1) and (2), intersite nonlinear terms
(accounting for an anharmonism of the intersite couplings) may also be added
to the model. The final results presented below in sections 3 and 4 are not
essentially altered by the intersite nonlinearities.

Linearization of Eqs. (1) and (2) and substitution of $x_{n},y_{n}\sim \exp
(ikn-i\omega t)$ yields two disjoint branches of the dispersion relation $%
\omega (k)$ for the two components: 
\end{mathletters}
\begin{equation}
\omega _{x,y}^{2}(k)=\Omega _{x,y}^{2}+4\kappa _{x,y}\sin ^{2}~(k/2)\,.
\label{dispersion}
\end{equation}
Note that, in principle, the coupling constants $\kappa _{x,y}$ may be both
positive and negative, corresponding, respectively, to the intersite
repulsion and attraction (in the case of the chain of dipoles, the repulsion
and attraction correspond, respectively, to the dipole components transverse
and longitudinal with respect to the chain, so that at least one coupling is
always positive, because at least one component is transverse). In any case,
we assume that both dispersion branches (\ref{dispersion}) are stable, which
requires $|\kappa |<\Omega ^{2}$, in the case when the coupling is negative.

We are looking for a case in which the dispersion branches generated by Eqs.
(1) and (2) correspond, respectively, to FH and SH. Then, the matching
conditions necessary for the existence of QS read that, at some point $%
k=k_{1}$, 
\begin{mathletters}
\begin{equation}
\omega _{y}(2k_{1}+Q)=2\omega _{x}(k_{1})+\Delta \omega ,  \label{match01}
\end{equation}
and 
\begin{equation}
v_{x}(k_{1})=v_{y}(2k_{1}+Q),  \label{match02}
\end{equation}
where $v_{x,y}(k)=d\omega _{x,y}/dk$ are the group velocities of the
respective modes, $\Delta \omega $ is a small frequency mismatch, $|\Delta
\omega |\ll \omega _{x,y}$, and $Q$ is a vector of the reciprocal lattice:
at $Q=0$ and $Q=\pm 2\pi $, one has {\em normal} and {\it umklapp}
processes, respectively. The presence of $Q$ in the matching conditions is a
specific feature of the discrete system.

First we concentrate on the exact matching conditions, when $\Delta \omega =0
$. After a straightforward algebra, these conditions yield 
\end{mathletters}
\begin{equation}
\cos k_{1}=\kappa _{x}/\kappa _{y}\,,  \label{k}
\end{equation}
hence the matching demands that $(\kappa _{x}/\kappa _{y})^{2}\leq 1$, i.e.,
the fundamental harmonic must be excited in a sublattice with a weaker
intersite coupling. With the wavenumber being fixed by Eq. (\ref{k}),
another final condition generated by the matching equations (\ref{match01})
and (\ref{match02}) is a constraint imposed on the model's parameters, 
\begin{equation}
4\Omega _{x}^{2}-\Omega _{y}^{2}=4\kappa _{y}^{-1}\left( \kappa _{y}-\kappa
_{x}\right) ^{2}\,.  \label{condition}
\end{equation}
Of course, QS are possible not only when this constraint is satisfied
exactly, but also in a vicinity of the corresponding surface in the system's
parameter space.

At the point where the matching conditions are exactly satisfied as
described above, the FH frequency can be easily found as $\omega _{1}\equiv
\omega _{x}(k_{1})$, 
\begin{equation}
\omega _{1}^{2}=\Omega _{x}^{2}+2\kappa _{x}\kappa _{y}^{-1}\left( \kappa
_{y}-\kappa _{x}\right) \,,  \label{fundamental}
\end{equation}
and the SH frequency is exactly $2\omega _{1}$.

Eqs. (\ref{k}) and (\ref{condition}) are the main results obtained from the
analysis of the model's linear part. It is interesting to note that the
nontrivial solution to the matching equations is specific for the discrete
model considered. Indeed, straightforward consideration of its continuum
counterpart (which can be formally obtained from Eqs. (\ref{dispersion}) in
the limit of infinitesimally small $k$) easily shows that the matching
conditions impose not one (cf. Eq. (\ref{condition})) but {\em two}
constraints on the model's parameters, viz., $\kappa_x =\kappa_y$, and $%
\Omega_y = 2\Omega_x$. If these two conditions are met, the SHG resonance in
the continuum model becomes degenerate, holding at all the values of $k$.
Thus, the case when the SHG resonance is possible in the continuum model is
quite trivial.

The model (1), (2) assumes no linear coupling between the two sublattices.
However, if a weak linear coupling between them is present, it may be
treated as a small perturbation. Analysis shows (we do not display here
boring technical details) that a weak linear coupling between $x_{n}$ and $%
y_{n}$ only slightly changes the relations (\ref{k}) and (\ref{condition}).
Thus, these results are structurally stable and can be extended to more
complicated models. This, in particular, means that the QS solitons may also
be generated by the SHG matching in molecular systems described by two
linearly coupled one-dimensional lattices (see, e.g., Refs. \cite{VoSa}).

\section{Nonlinear Analysis}

It is customary to seek for QS by means of multiscale expansions, which in
our case must be implemented in the discrete model. To this end, we
introduce a formal small parameter $\mu $ ($\mu \ll 1$), and look for a
solution to Eqs. (\ref{eq1a}) as (\ref{eq1b}) in the form 
\begin{mathletters}
\beq
\label{anz1}
x_{n}=\mu ^{2}A_{x}(t_{1},...;\xi _{1},...)e^{ik_{1}n-i\omega _{1}t}+{\rm %
c.c.}\,
\enq
\beq
\label{anz2}
y_{n}=\mu ^{2}A_{y}(t_{1},...;\xi _{1},...)e^{2\left( ik_{1}n-i\omega
_{1}t\right) }+{\rm c.c.}\,,
\enq
where we have introduced a set of spatial and temporal variables $%
t_{j}\equiv \mu ^{j}t$ and $\xi _{j}\equiv \mu ^{j}n$ ($j=1,2,...$) which
are regarded as independent and {\em continuous} ones, $A_{x,y}$ are
amplitudes which, in principle, depend on all the variables, and ${\rm c.c.}$
stands for the complex conjugate. It is assumed that the carrier wave number
and frequency, $k$ and $\omega $, exactly satisfy the SHG matching
conditions (\ref{match01}), (\ref{match02}).

We notice that the above substitution implies an infinite lattice, which,
however is not a limitation for the analysis: essentially the same results
can be obtained for a finite lattice with a large number of sites, see e.g. 
\cite{VK1,VK2}.

Substituting the ansatz (\ref{anz1}), (\ref{anz2}) into Eqs. (\ref{eq1a})
and (\ref{eq1b}), and collecting all the terms of the same order with
respect to $\mu $, at the lowest (second) order we recover the dispersion
relation, and then at the third order we obtain that $A_{x,y}$ depend on $%
t_{1}$ and $\xi _{1}$ only through $R=\xi _{1}-vt_{1}$, $v$ being the group
velocity of the modes. Finally, at the forth order with respect to the small
parameter $\mu $ we arrive at a system of continuous equations, 
\end{mathletters}
\begin{mathletters}
\begin{equation}
i\frac{\partial A_{x}}{\partial t_{2}}+\frac{\omega _{1}^{\prime \prime }}{2}%
\frac{\partial ^{2}A_{x}}{\partial R^{2}}+\chi _{x}\bar{A}%
_{x}A_{y}e^{-i\delta \omega \cdot t_{2}}=0  \label{eq2a}
\end{equation}
\begin{equation}
i\frac{\partial A_{y}}{\partial t_{2}}+\frac{\omega _{2}^{\prime \prime }}{2}%
\frac{\partial ^{2}A_{y}}{\partial R^{2}}+\chi _{y}A_{x}^{2}e^{i\delta
\omega \cdot t_{2}}=0  \label{eq2b}
\end{equation}
where $\omega _{1}^{\prime \prime }=d^{2}\omega _{x}/dk^{2}|_{k=k_{1}}$ and $%
\omega _{2}^{\prime \prime }=d^{2}\omega _{x}/dk^{2}|_{k=2k_{2}+Q}$ are the
group velocity dispersions of the first and second modes, $\chi _{x}\equiv
a_{xx}/(2\omega _{1})$, $\chi _{y}\equiv b_{xx}/(4\omega _{1})$,  it is
assumed that $\Delta \omega =\mu ^{2}\delta \omega $ with $\delta \omega
=O(\omega _{1})$, so that $\Delta \omega \cdot t\equiv \delta \omega \cdot
t_{2}$, the overbar stands for the complex conjugation, and dependence of $%
A_{x,y}$ on $\xi _{2}$ is neglected.

Eqs. (\ref{eq2a}) and (\ref{eq2b}) are identical to general equations
describing SHG in the {\it temporal domain} in dispersive optical media
(see, e.g., \cite{BurKiv}), $\delta \omega $ playing the role of the
standard mismatch parameter. Thus, these equations give rise to stable
solitons of the well-known form.

\section{Multidimensional Case}

Coming back to the interpretation of the underlying model (1), (2) as that
describing interaction of two-component dipoles pinned at the lattice sites,
it is straightforward to extend it to the two- and three-dimensional (2D and
3D) cases. Because the host medium is assumed anisotropic, in the most
general case all the coupling constants may be different. Thus, in the 2D
case, for instance, the dynamical variable are $x_{nm}$ and $y_{nm}$, the
``discrete Laplacian'' in Eq. (1) being replaced by 
\end{mathletters}
\begin{eqnarray}
&&\kappa _{x}^{{\rm (l)}}\left( x_{n+1,m}-2x_{nm}+x_{n-1,m}\right)  
\nonumber \\
&&+\kappa _{x}^{{\rm (t)}}\left( x_{n,m+1}-2x_{nm}+x_{n,m-1}\right) \,,
\label{y}
\end{eqnarray}
where $\kappa _{x}^{{\rm (l)}}$ and $\kappa _{x}^{{\rm (t)}}$ are the
coupling constants in the longitudinal and transverse directions. Eq. (2) is
to be modified in a similar way. Note that solitary waves in the
corresponding monoatomic lattice in the strongly anisotropic limit, $\kappa
_{x}^{{\rm (t)}}\ll \kappa _{x}^{{\rm (l)}}$, have been considered in Ref. 
\cite{DEWZ}.

In the 2D case, the matching conditions read: 
\begin{equation}
\omega _{y}(2{\bf k}_{1}+{\bf Q})=2\omega _{x}({\bf k}_{1})+\Delta \omega ,
\label{match11}
\end{equation}
and 
\begin{equation}
{\bf v}_{x}({\bf k}_{1})={\bf v}_{y}(2{\bf k}_{1}+{\bf Q})  \label{match12}
\end{equation}
where ${\bf v}_{x,y}({\bf k})=d\omega _{x,y}({\bf k})/d{\bf k}$, the bold
letters standing for 2D vectors having longitudinal and transversal
components; for instance, ${\bf k}\equiv (k^{{\rm (l)}},k^{{\rm (t)}})$.

In the 2D version of the present model, the exact matching conditions can be
cast into a form 
\[
\cos k_{1}^{{\rm (l)}}=\frac{\kappa _{x}^{{\rm (l)}}}{\kappa _{y}^{{\rm (l)}}%
},\,\,\,\,\,\,\,\cos k_{1}^{{\rm (t)}}=\frac{\kappa _{x}^{{\rm (t)}}}{\kappa
_{y}^{{\rm (t)}}}\,.
\]
More sophisticated schemes with a 2D configuration of the FH and SH
wavevectors are possible too, but we do not consider them here.

The derivation of the equations for the envelope functions, outlined above
for the 1D case, can be easily extended to the 2D case. As a result, at the
second order in the formal small parameter $\mu $ defined above, the result
is 
\[
\frac{\partial A_{x,y}}{\partial t_{1}}+({\bf v}\nabla _{1})A_{x,y}=0,
\]
where $\nabla _{j}\equiv \partial /\partial {\bf \xi }_{j}$ and $\xi _{j}$
is now a 2D vector, $\xi _{j}\equiv (\xi _{j}^{{\rm (l)}},\xi _{j}^{{\rm (t)}%
})$. From here on we conclude that the envelope amplitudes are functions of
the traveling-wave argument ${\bf R}={\bf \xi }_{1}-{\bf v}t_{1}$. Finally,
at the third order in $\mu $, we obtain a system 
\begin{mathletters}
\begin{eqnarray}
&&i\frac{\partial A_{x}}{\partial t_{2}}+\frac{1}{2}\sum_{\alpha _{1},\alpha
_{2}={\rm l},{\rm t}}\omega _{x,\alpha _{1},\alpha _{2}}\frac{\partial
^{2}A_{x}}{\partial R^{\alpha _{1}}\partial R^{\alpha _{2}}}  \nonumber \\
+\chi _{x}\bar{A}_{x}A_{y}e^{-i\delta \omega t_{2}} &=&0,  \label{umultiD}
\end{eqnarray}
\begin{eqnarray}
&&i\frac{\partial A_{y}}{\partial t_{2}}+\frac{1}{2}\sum_{\alpha _{1},\alpha
_{2}={\rm l},{\rm t}}\omega _{y,\alpha _{1},\alpha _{2}}\frac{\partial
^{2}A_{y}}{\partial R^{\alpha _{1}}\partial R^{\alpha _{1}}}  \nonumber \\
+\chi _{y}A_{x}^{2}e^{i\delta \omega t_{2}} &=&0,  \label{vmultiD}
\end{eqnarray}
where $\omega _{\alpha ,\alpha _{1},\alpha _{2}}=\partial ^{2}\omega
_{\alpha }/\partial k^{\alpha _{1}}\partial k^{\alpha _{2}}|_{{\bf k}={\bf k}%
_{1}}$ ($\alpha =x,y$ and $\alpha _{1},\alpha _{2}={\rm (l),(t)}$) is the
group-velocity-dispersion tensor.

The generalization for the 3D case is straightforward. It is necessary to
stress that Eqs. (\ref{umultiD}) and (\ref{vmultiD}) are more general than
the multidimensional SHG equations known in nonlinear optics \cite{bullet},
as they are fully anisotropic and contain {\em mixed} second derivatives in
the variables $R_{j}$. Accordingly, these equations can give rise to more
general multidimensional solitons than those known in the optical models 
\cite{bullet}. However, a detailed consideration of these more general
multidimensional solutions is beyond the scope of this work.

\section{Conclusion}

In this work, we have proposed a two-component lattice model, in its one-
and multidimensional versions, which may satisfy the matching conditions
necessary for the generation of solitons through the second-harmonic
generation. The model describes an array of linearly coupled two-component
dipoles in an anisotropic nonlinear host medium. Unlike this discrete
system, its continuum counterpart meets the matching conditions only in a
trivial degenerate situation. We have derived nonlinear evolution equations
for slowly varying envelope functions of the resonantly coupled fundamental
and second harmonics. In the one-dimensional case, these equations naturally
prove to be tantamount to the standard system well-known in nonlinear
optics, while in the multidimensional case the system is more general than
the optical one, thanks to the anisotropy of the underlying lattice model.

The analysis developed in this work can be extended in different directions.
In particular, one can consider more general three-wave resonant
interactions, which are also known to give rise to solitons in optics \cite
{3W}. In the context of condensed-matter lattice models, resonant three-wave
interactions are, e.g., known in the model of the electron-phonon system of
the Davydov's type \cite{VK}, but an analysis shows that the full set of the
matching conditions in that model can only be met in a limiting case of
equal maximum electron and phonon frequencies, which is not physically
realistic. A fully matched three-wave resonance is possible in other models,
but a detailed consideration of this problem will be presented elsewhere.

Note also that, in the model considered above, the coupling of the two
components is purely nonlinear. Adding a linear coupling results in a more
complicate dynamics, when QS are excited together with so-called
accompanying modes (i.e., nonresonantly excited ones).

Lastly, dynamics of real crystallic lattices is usually characterized by
conspicuous dissipative losses. The simplest way to compensate the losses is
by an external ac drive applied to the edge of a sample, the drive's
frequency being equal to the FH frequency. However, detailed analysis of the
loss compensation by an external pump should be a subject of another work.

\section{ACKNOWLEDGEMENTS}

B.A.M. appreciates hospitality of the Department of Physics at the
Universidade da Madeira (Funchal, Portugal). V.V.K. acknowledges support
from FEDER and Program PRAXIS XXI, grant No. PRAXIS/2/2.1/FIS/176/94.

\end{mathletters}

\end{document}